# "World Wine Exports: What Determined the Success of the 'New World' Wine Producers?"

**Osiris Jorge Parcero**

Research Affiliate CRIEFF, University of St. Andrews

**Emiliano Villanueva**

Research Affiliate AUIV, International Organisation of Vine and Wine

July 20, 2011

## Abstract

By using an econometric approach this paper looks at the evolution of the world wine industry in the period 1961-2005. A particular stylized fact is the appearance of non-traditional producing and exporting countries of wine from the beginning of the nineties. We show that the success of these new producing and exporting countries can be explained by the importance of the demand from non-producing countries with little or no tradition of wine consumption, relative to the world demand. This stylized fact is consistent with a testable implication of the switching cost literature and to the best of our knowledge this is the first time that this implication is tested.

# 1. Introduction

Traditionally, wine producing countries consumed their own wine and hence wine exports, which mainly occurred between neighboring countries, were of only 10% of the liters of wine sold in the global market in the period 1961-1965. However, the proportion of wine traded internationally has grown substantially in recent decades, reaching 30% of the global wine sales for the period 2001-2005. International trade in wine has increased by almost 200% from 1961 to 2005 and, even though there was a slight drop in per capita wine consumption overall, the wine has become a product highly traded internationally.

This significant increase in wine exports began in the traditional European wine producers in the seventies, followed by new producing and exporting countries; i.e. the U.S. and Australia since the eighties and Chile, Argentina and South Africa since the nineties (Villanueva, 2011). Until the late '80s the export of wine has been an almost exclusive European business (95% of the total wine exports in 1980 were European), but since then, the proportion of wine exported from Europe began to decline, dropping to less than 70% in 2005. This market-share was taken away by those new producing and exporting countries.

The wine industry in the traditional European wine producers was generally characterized by a production-consumption system based upon a fairly stable business sector in which cooperatives (the majority) coexisted with companies (mostly familiar and small and medium-sized). "Table wine" had a large participation in the wine production, while the quality wines, based on the "Denomination of Origin" system, represented a more limited share. However, at the beginning of the 21st century the scenario is substantially different. While the production of table wine goes through a long and deep crisis, the wines produced under the denominations of origin method gain considerably importance.

A historical anecdote that perhaps shows the start of this deep and clear process of change in the global wine industry is the legendary professional tasting held in Paris in May 1976, known as the "Judgment of Paris". In the tasting a couple of unknown Californian wines beat some of the most famous French ones. At that time, the possibility of an export boom being led by what later became known as the new wine producing and exporting countries was unthinkable. However, the evolution of wine exports since then seems to prove otherwise. As an example we can mention that Australian wine exports in 1961 were almost zero and at the end of 2005 they reached 10% of the global wine exports; Chilean wine exports in 1961 were nil, but in 2005 Chile exported nearly 5% of the global wine exports.

Simultaneously with the emergence of these new producing and exporting countries, there was the appearance of a significant increase in demand of wine from new buyers; i.e., from non-producing countries with little or no tradition of wine consumption. The consumption profile in these new consumer countries was qualitatively different from the one in the traditional consumer countries. The preferences of these new consumers were likely to be less subject to tradition and / or reputation of certain areas and established brands and therefore more likely to try new wines.

In the previous paragraphs we made the point that the world wine industry is characterized by the existence of three different groups of countries, i.e. traditional (or established) producers, new producers and new buyers. Moreover, we made it clear that one stylized fact is that the appearance of these new buyers occurs at a stage where the traditional producers were already well established. A second stylized fact is the simultaneous emergence and success of the new wine producers who rapidly gained market share from the traditional ones.

The paper shows that these two stylized facts are consistent with a testable implication posited in the switching cost literature, a sub-field of Industrial Economics. This approach analyzes oligopolistic competition in the presence of consumers' switching cost, where incumbent firms face the threat of entrance of new competitors. In our context this implication can be read as follows: the success of the new producing and exporting countries can be explained by the importance of the demand from non-producing countries with little or no tradition of wine consumption, relative to the world demand.

By combining a historical perspective and an econometric approach, this paper is an important contribution in the direction of getting a better understanding of the evolution of the world wine exports. Moreover, it not only explains an important world wine industry stylized fact but, to the best of our knowledge, this is the first time that the aforementioned switching costs' implication has been tested.

The structure of the paper is as follows. Section 2 explores the related literature and presents the hypothesis. Section 3 introduces the data and proposes the empirical model. Section 4 shows the results. Section 5 concludes.

## 2. Literature Review and Hypothesis

Our paper is connected to two separate branches of literature. The first one looks at firms' exports propensity and the second one is related to how the existence of "switching costs" may favor the entrance of new competitors in an industry. The former has mainly concentrated on the determinants of exports propensity at firm level, where the firms may belong to the same industry or different industries and are located in one country or a small group of countries. Some of the typical determinants that this literature looks at are the plant and firm size, technology, whether or not the firm has foreign owners, market structure, R&D and human capital. In the econometric specification this literature uses the firm´s export propensity as the dependent variable, which is defined as the firm's exports divided by its total sales. This literature has mainly focused on industrialized countries and among the most recent papers are Wagner (1995), Wakelin (1998), Lefebvre et al. (1998), Nassimbeni (2001), Sterlacchini (2001), Van Dijk (2002), Ropern et al. (2006), Greenaway and Kneller (2007) and Kneller at al. (2008) to only mention a few.

Our paper differentiates from this literature in that it concentrates on a particular industry (instead of an aggregate of them), but where the units under consideration are countries instead of firms, making our study more aggregate in this way. The disadvantage of using this more aggregate data is counterbalanced by the fact that we look at the evolution of one particular industry in the whole world and throughout a very long period of time.

The second branch of literature is a sub-field of Industrial Economics that analyzes oligopolistic competition in the presence of consumers' switching cost, where competition for established buyers is continually intermingled with competition for new ones. Of a particular interest to our paper is the 'open' oligopoly case, where incumbent firms face the threat of entrance of new competitors.

This literature predicts that when the new-buyers' demand grows slowly relatively to the one coming from the old (locked-in) ones, the entrance of new firms into a market with switching costs may be very hard (Schmalensee 1982, Farrell 1986, Klemperer 1987 and 1995). However, when the relative importance of the demand coming from new buyers is high (a situation usually associated with a rapid market growth), the entry of new firms may be possible. On the one hand, this may be the case if the new entrants' lower cost of production gives them a competitive advantage which, though not enough to steal the locked-in buyers from the incumbent, becomes significant at the time of stealing the new buyers. On the other hand, and even in the absence of this competitive advantage, incumbent firms may find it difficult to charge high prices to exploit their old customers and at the same time charge low prices to compete with new entrant firms for the new customers, i.e., price discrimination is difficult (Klemperer 1987, Farrell and Shapiro 1988, Beggs and Klemperer 1989, Gabszewicz et al.1992, Wang and Wen 1998, Farrell and Klemperer 2007). That is, if the willingness to exploit the old buyers were strong enough the incumbent firms would set a relatively high price and by doing so would lose many of the new buyers to the new entrants or less established firms. In other words, the appearance of the new (not locked-in) buyers makes the entrance of new firms into the market easier as well as helps firms with small market shares to increase them.

In the words of Farrell and Klemperer (2007):
> "… if firms cannot discriminate between old and new consumers, then the "fat cat" effect may make small scale entry very easy: incumbent firms' desire to extract profits from their old customers creates a price umbrella under which entrants can profitably win new unattached (or low switching cost) customers. And even after entry has occurred, the erstwhile incumbent(s) will continue to charge higher prices than the entrant, and lose market share to the entrant, so long as they remain "fatter" firms with more old consumers to exploit."

Even though the unit of analysis is different, the world wine export market has some of the features that are typical in the switching cost model just described – i.e., it also has traditional (or established) producers, new producers and new buyers. In the wine literature it is common to divide the world into the Old Wine World (OW) and the New Wine World (NW). The OW is composed by established producers (France, Italy, Spain, Portugal, Germany, Austria, Switzerland, Belgium, Greece, Bulgaria, Hungary and Romania) while the NW is composed by new producers (United States, Australia, New Zealand, South Africa, Canada, Argentina, Chile, Brazil, Mexico and Uruguay)[1]. As can be seen in Chart 1a the OW has a substantially higher exports' share than the NW throughout the period 1961-2005 and the NW's share in world wine exports is nil until the beginning of the 80's. Thus, in terms of the switching cost framework just explained, the OW's countries will be considered as the established exporters while the NW's countries will be the new entrants. Moreover, as a whole, the countries selected as part of the Old and New World Wine account for 90.9% and 76.5% of global production and consumption of wine respectively ("Global Wine Statistical Compendium, 1961-2005", Adelaide, hereafter GWSC).

**[Chart 1 here]**

All the countries not included in the NW and the OW (hereafter Rest of the Word, RW) are countries with little tradition of wine consumption and practically inexistent tradition of wine production but, as can be seen in Chart 1b, have significantly increased their imports relatively to the world's total imports of wine throughout the period 1961-2005. Among these countries stand out United Kingdom, Denmark, Switzerland, Holland, Russia, Belgium, Japan, China and India, which are the biggest importers inside this group. Thus, for the purpose of our research, the RW can be considered as what it is defined as "new buyers" in the mentioned switching costs literature.

It is clear that the aforementioned switching cost models were developed to explain markets where the actors were firms and in our wine industry analysis the data only allow us to identify countries as the unit of production and export. That is, these models were designed for a case where the switching costs are firm's specific and where these firms individually take advantage of them by setting higher prices than their competitors. On the contrary, in the case of countries the appearance of switching costs and setting of prices is a much more subtle thing. History, national pride, as well as many regulations and/or coordinated actions from the OW producers (at local, regional and national levels) and their governments have resulted in the emergence of switching costs. For instance, the Denominations of Origin significantly increased the market segmentation and the belief that products from some specific regions were of substantially higher value, which resulted in higher prices. Coordinated advertisement and the history of more than one thousand years have increased the value of their wines in people's imaginary, relatively to wines from countries that are not traditional producers, resulting in a higher product loyalty[2]. For instance, even though for many years the quality of many sparkling wines produced around the world got substantially closer to that of the Champagne, the latter was and still is sold for a much higher price than the former ones. Moreover, when talking to a French (Italian, Spanish, Portuguese, etc.) person, one could easily realize that he would consider many of his country's wines as superior to similar quality wines produced somewhere else. However, people from many other nationalities, for instance British, whom for a long time have been exposed to wines imported from different parts of the world, would be less biased.

---

[1] These designations of Old and New Worlds are very common in the wine literature; they can be found in countless newspaper articles, publications and academic studies of different fields: agronomy, oenology business, economics and law. Some studies that use them are Anderson (2001), Anderson et al. (2003), Campbell and Guibert (2006), Green et al. (2003), Cholette et al. (2005), Duncan and Greenaway (2008) and Simpson (2009). A detailed analysis of the evolution of these concepts can be found in Villanueva (2011).

From the previous paragraph argument it is reasonable to think that an increase in demand coming from the OW has relatively more of the locked-in type of buyers than an increase in demand coming from the RW. The same argument would also apply to an increase in demand coming from the NW, which could be more locked-in to the NW producers. However, it is highly expected that this lock-in effect is smaller in the NW than in the OW; the markets of the former were much smaller and in some cases insignificant.

The previous discussion leads to the following claim. If the increase in the world wine demand comes from the increase in the demand of costumers in the OW (NW), this would result in a higher demand of wines produced in the OW (NW) relatively to those produced in the NW (OW). On the contrary, this bias will not be seen if the increase in demand originates in the RW. Thus, if the switching cost is strong enough we would expect the demand from the RW to benefit relatively more the producers of the NW than the ones in the OW.

---

[2] In a different context this has been posit by Howard and Sheth (1969). Their research shows that households may routinize their brand purchases by buying the same brand repeatedly over time, something that they called state dependence or inertia.

One important contribution of our paper is to the empirical literature on switching costs. This literature is much smaller and more recent than the theoretical one. First, few scholars have attempted to directly measure the switching cost. For instance, after controlling for other factors that may have influenced the vendor-buyer match, Greenstein (1993) suggests that switching costs were an important source of incumbent advantage in the US federal procurement of commercial mainframe computer systems during the seventies. Similarly, Shum (2004) analyzes panel data on breakfast cereal purchases and measures the average switching cost incurred by households moving to other brands.

Second, some studies have looked at the importance of switching cost as a source of incumbent advantage. Looking at prices that airlines charge to different passengers on the same route, Borenstein and Rose (1994) find that the pattern of observed price dispersion cannot be only explained by cost differences. Their finding is an indirect one because the importance of switching cost as a source of incumbent advantage is obtained by observing a dispersion increase on routes subject to more competition. They conclude that this is consistent with discrimination based on customers' willingness to switch to alternative airlines. Borenstein (1991) analysis of gasoline retailing in the United States points in the direction that leaded gasoline is priced higher than unleaded because, since there are fewer stations selling it, buyers of leaded gasoline face higher costs of switching from one station to another.

Third, an extensive empirical marketing literature focus on brand loyalty (or "state dependence") which often reflects, or has equivalent effects to, switching costs; Seetharaman et al. (1999) summarize this literature.

To the best of our knowledge, the empirical literature of switching costs has not look at the issue that the arrival of new (unattached) buyers would lead to the appearance and / or success of new competitors. Thus, this becomes our contribution to the switching cost empirical literature.

Our hypothesis seems to be very well supported by the stylized fact in the period 1981-85 to 2001-05 (see Chart 1). That is, the increase in the RW imports' share concurrently occurs with a fall (rise) in the OW (NW) exports' share. However and quite on the contrary, in the period 1961-65 to 1971-75 the increase in the RW imports' share simultaneously occurs with a rise (no variation) in the OW (NW) exports' share. Thus, a quick glance at Chart 1 does not seem to suggest that the support for our hypothesis is unequivocal and obvious.

It would be in our advantage to argue that the data in the period 1961-65 to 1971-75 does not really contradict our hypothesis for the following reason. The claim that during this period the NW was in fact a "new entrant" to the exports market only in a potential way because its exports' share was practically nil (around 1.2%). It may have not been able to exploit this potential because of the lack of economies of scale at the industry level. Moreover, in view of the very low levels of world wine exports at the time, we could claim that the OW may have played the entrant's role in the export market instead; its exports were increasing as a result of a surge in international trade that accompanied the postwar. Thus, we may be willing to round up by saying that a fairer empirical testing of our hypothesis should be done in a period where the incumbent is already established in the market, which only occurs at a later stage. Nevertheless, in our econometric analysis of the following section we look at the whole period 1961-65 to 2001-05. That is, the period in which we test our hypothesis includes those initial years in which, at first glance, it seems less likely that it would hold.

Moreover, there is a historical issue that is strongly affecting the shares observed in Chart 1a. During the first five-year periods the OW exports' share can be considered as substantially underestimated because there were large exports by the North of Africa (i.e. part of RW). North Africa wine exports were mainly from Algeria and Tunisia, political territories of France at that time. In Chart 2 we have done the exercise of adding the North African exports to that of the OW and we can clearly see that by doing so our hypothesis seems to become supported by the data even in the period 1961-65 to 1971-75.

**[Chart 2 here]**

As it is clear by now, our main hypothesis is to show that the appearance of new buyers relatively favors new producers. However, we are aware that there are three other potential reasons that may explain the NW success relative to the OW. All these reasons, in one way or another, are obviously interrelated and can influence each other. The first one is the business and productive model implemented by the NW, which corresponds to a mass production of varietal wines, with quality consistency, obtained by large industrial enterprises with significant economies of scale. By contrast and as previously mentioned, the productive and business model implemented by the OW was characterized by a business network of cooperatives and enterprises (mostly familiar and of small and medium size). The second one is the commercial and marketing strategy implemented by the NW, leading to better consumer information of the product characteristics; i.e. wines are sold under brands with strong investments in marketing and advertising campaigns. By contrast, the commercial tradition of the OW involved an intricate system of denominations of origin, varietal and geographic areas that were difficult to understand by a novice consumer. The last potential reason is the emergence of strong governmental support in the NW during the '90s and the beginning of this century. This support, known in the specialized literature as the "National Brand Plans", had the primary aim of improving the wine export performance in the long term[3].

## 3. Data and econometric model

The empirical analysis presented here tests our main hypothesis – i.e. the appearance of the demand coming from the RW was an important determinant of the increase in the market share of the NW relative to the one of the OW. In order to do that we use a panel of 47 geographical units (i.e. 38 countries and 9 regions) and 9 periods but, for simplicity hereafter we will refer to the geographical units as "countries". The variable on wine exports and imports in both volume and value are obtained from the GWSC, which are available as a five-year average from 1961-65 to 2001-05.

The GWSC reports data for the countries that are the main wine producers and exporters of the world, but agglomerates in regions to those countries with very low production and/or exports. The countries and regions are: Argentina, Australia, Austria, Azerbaijan, Belgium-Luxembourg, Brazil, Bulgaria, Canada, Chile, China, Croatia, Denmark, Finland, France,

---

[3] Australia was the pioneer in the development of such national scope plans in June1996, launching its celebrated plan called "Strategy 2025". Its aim was to reach an export turnover of 4.5 billion Australian dollars by the year 2025. The resounding success of the plan meant that in 2005, 20 years earlier than planned, the target was achieved. The Australian plan was an inspiration for the national plans of the United States and Chile, followed soon after by Argentina and South Africa.

Georgia, Germany, Greece, Hungary, Ireland, Italy, Japan, Mexico, Middle East[4], Moldova, Netherlands, New Zealand, North Africa[5], Other Africa[6], Other Asia-Pacific[7], Other Central and Eastern Europe[8], Other Latin American and Caribbean[9], Other North and East Asia[10], Other West Europe[11], Portugal, Romania, Russia, South-East Asia[12], South Africa, Spain, Sweden, Switzerland, Turkey, Ukraine, United Kingdom of Great Britain, United States, Uruguay and Uzbekistan.

In order to test our hypothesis we run the following regression:

---

[4] Afghanistan, Jordan, Saudi Arabia, Bahrain, Kuwait, Syria, Gaza, Lebanon, UAE, Iran, Yemen, Iraq, Oman, Israel and Qatar.

[5] Algeria, Morocco, Egypt, Tunisia and Libya.

[6] Angola, Equatorial Guinea, Mali, Sao Tome and Principe, Benin, Eritrea, Mauritania, Senegal, Botswana, Ethiopia, Mauritius, Seychelles, Burkina Faso, Gabon, Mozambique, Sierra Leone, Burundi, Gambia, Somalia, Cameroon, Ghana, Sudan, Cape Verde, Guinea, Swaziland, Central African Republic, Guinea-Bissau, Namibia, Tanzania, Chad, Kenya, Niger, Uganda, Congo, Lesotho, Nigeria, Western Sahara, Congo-Brazzaville, Liberia, Rwanda, Zambia, Cote D'Ivoire, Madagascar, Reunion, Zimbabwe, Djibouti and Malawi.

[7] American Samoa, Kiribati, New Caledonia, Solomon Islands, Bangladesh, Maldives, Palau, Sri Lanka, Cook Islands, Marshall Islands, Pakistan, Tokelau, Fiji, Mongolia, Papua New Guinea, French Polynesia, Samoa, Vanuatu, India, Nepal, Wallis Islands and Tonga.

[8] Albania, Lithuania, Armenia, Macedonia, Belarus, Poland, Bosnia-Herzegovina, Serbia, Czech Republic, Slovakia, Estonia, Slovenia, Kazakhstan, Turkmenistan, Kyrgyzstan and Latvia.

[9] Anguilla, Peru, Antigua, Barbuda, Costa Rica, Guatemala, Puerto Rico, Bahamas, Cuba, Guyana, Saint Lucia, Barbados, Dominica, Haiti, Islands Saint Kitts and Nevis, Belize, Dominican Republic, Honduras, Bermuda, Ecuador, Jamaica, Saint-Pierre, Bolivia, El Salvador, Martinique, Suriname, British Virgin Islands, French Guyana, Nicaragua, Turks Islands, Cayman Islands, Grenada, Panama, Trinidad and Tobago, Colombia, Guadeloupe, Paraguay and Venezuela.

[10] Hong Kong, Korea, Macao, Taiwan and North Korea.

[11] Andorra, Greenland, Cyprus, Iceland, Faroe Islands, Liechtenstein, Finland, Malta, Gibraltar and Norway.

[12] Brunei, Myanmar, Cambodia, East Timor, Singapore, Indonesia, Thailand, Laos, Vietnam, and Malaysia.

$$EXP_{it} = \beta_0 + \beta_1 NIRW_t + \beta_2 OW_i + \beta_3 NW_i + \beta_4 NIRW\_x\_OW_{it} + \beta_5 NIRW\_x\_NW_{it} +$$
$$\beta_6 RMP_{it} + \beta_7 EU68\text{-}98_{it} + \beta_8 EURO99\text{-}05_{it} + \beta_9 CHRIST\_RULER_i +$$
$$\beta_{10} MUSLIM\_RULER_i + \beta_{11} DISTLAT3050_i + \beta_{12} AVERAGE\_TEMP_i + \beta_{13} QUINQ61\text{-}65_t +$$
$$\beta_{14} QUINQ61\text{-}65\_x\_OW_{it} + \beta_{15} QUINQ61\text{-}65\_x\_RW_{it} + \alpha_i + \mu_{it} \qquad (1)$$

where the subscript i and t indicate the geographical unit and the period respectively while $\alpha_i$ and $\mu_{it}$ are the random effects and the error term respectively. The choice of a random effect model is justified because, as it is clear in the equation, we have 6 time-invariant variables.

The dependent variable is "exports' share", which is defined as the country's exports divided by the total global exports. In order to make our results as robust as possible the regression analysis will be done with two different specifications, one with the exports and imports measured in volume and the other measured in value. So whenever we mention exports or imports we will not specify whether they are expressed in volume or value until we discuss and describe the results in the following section. The independent variables and their expected effects are described below.

"*OW*" is a dummy variable that takes the value 1 if the country belongs to the "Old World" and 0 otherwise. "*NW*" is a dummy variable that takes the value 1 if the country belongs to the "New World" and 0 otherwise. The variable $NIRW_t$ is defined as follows

$$NIRW_t = \frac{\sum_{h \in RW} NI_{ht}}{\sum_{k=1}^{47} GI_{kt}} \qquad (2)$$

It shows, for each period t, the sum of the net wine imports (*NI*) for all countries belonging to the *RW* divided by the world´s wine gross imports (*GI*)[13]. Along with two interactive effects defined later, this is the most critical variable in terms of testing our hypothesis. The variable $NIRW_t$ indicates the relative importance of demand from the *RW*, or simply the "relative demand of *RW*." Then, the coefficient $\beta_1$ in (1) shows the effect that the "relative demand of *RW*" has on the exports' share of each country belonging to *RW*.

---

[13] The reason why we use this ratio is as follows: the Net Imports (NI) of the Rest of the World (RW) is the best indicator of the RW's demand for the wine produced in each of the "wine´s worlds". However, since the net imports of the world total add up to zero, in the denominator of (2) we chose to include the world's wine gross imports (GI) instead.

The variable $NIRW\_x\_OW_{it}$ is the product of the variables $NIRW_t$ and $OW_i$. As a result, the addition of $\beta_1 + \beta_4$ indicates the effect that the relative demand of the *RW* has on the exports' share of each country belonging to the *OW*. Similarly, the variable $NIRW\_x\_NW_{it}$ is the product of the variables $NIRW_t$ and $NW_i$, and the addition of $\beta_1 + \beta_5$ indicates the effect that the relative demand of the *RW* has on the exports' share of each country belonging to the *NW*.

In the previous three paragraphs we have described the effect that $NIRW_t$ has on the exports' share for each of the countries belonging to each of the 'wine worlds'. Clearly, the effect that $NIRW_t$ has on the aggregate exports' share of each of the 'wine worlds' is the addition of the effect on each of its member countries. Moreover, the addition of the aggregate exports' share of the three wine worlds must add up to zero. That is,

$$\beta_1 n_{rw} + (\beta_1 + \beta_4)n_{ow} + (\beta_1 + \beta_5)n_{nw} = 0 \qquad (3)$$

where $n_{rw}$, $n_{ow}$ and $n_{nw}$ are the number of countries in the *RW*, the *OW* and the *NW* respectively. Expression (3) is a restriction that has to be imposed in the regression.

The variable *RMP* is the Real Market Potential; this concept is adapted from Head and Mayer (2006) and measures a country's potential access to the demand from all the other countries. The *RMP* for country $i$ is calculated as the weighted average of the GDP per capita of each country $j$, for $j \neq i$, where the weights are the inverse of the distance between country $i$ and each of the countries $j$. It is expected that the exports' share of a country would be positively related to *RMP*.

To measure the impact upon wine trade produced by the considerable extensions of the European agreement, two dummy variables are added, as it is standard in the trade literature, e.g. Serrano and Pinilla (2009). The variable $EURO99\text{-}05_{it}$ is a "dummy" that takes the value 1 from 1999 and until 2005 for all members of that Monetary Union and a value of zero otherwise. Similarly, the variable $EU68\text{-}98_{it}$ is a "dummy" that takes the value 1 for the years between 1968 and 1998 for all member countries of the European Union and a value of zero otherwise. 1968 and 1999 are the years of the creation of the European Union and the "Euro Zone" respectively.

Given that alcoholic beverages are not allowed in Islamic countries and that the wine has a long tradition in the Christian religion, we have added the following two variables. The variable $CHRIST\_RULER_i$ is a categorical variable with a value of 1 for any country having more than 50% of Christian population and that is mostly governed by the laws and customs of this religion, a value of 0.5 for the Middle East[14] and a value of zero otherwise. Similarly, the variable $MUSLIM\_RULER_i$ is a categorical variable where 1 indicates that a country has more than 50% of Muslim population and is governed by the laws and customs of this religion, 0.5 for Middle East and 0 otherwise. The data used to produce these two variables and are the percentages of religious militancy in each country obtained from "The CIA World Factbook" (2009).

The vine grows successfully between 30° and 50° latitude north and south where its distinctive natural annual cycle can be accommodated. This cycle involves harsh winters, with temperatures less than 0°C, but where the daily average should reach 10°C in the spring before the buds grow to 20°C in summer for flowering. The variable $DISTLAT3050_i$ a) indicates the distance between the average latitude of each country and the latitude 50°, if the country has a higher average latitude than 50°; b) indicates the distance between latitude 30° and the average latitude of each country, if the country has a lower average latitude than 30°; and c) is set to zero for all countries whose average latitude is between 30° and 50°. The variable $AVERAGE\_TEMP_i$ measures the average annual temperature (C°): a) the main wine region, if the country is a wine producer and b) the capital of the country if the country is not a wine producer. The data were obtained from "The CIA World Factbook" (2009).

The period 1961-65 is a special one because the large exports of Algeria (North Africa) positively impacted the exports of *RW*, but Algeria was in fact a political jurisdiction of France. Thus, we created the dummy $QUINQ61\text{-}65_{it}$ that takes the value 1 for the period 1961-65 and 0 otherwise. Moreover, given that Algeria belongs to the *RW* and France to the *OW* we decided to create the variable $QUINQ61\text{-}65\_x\_OW_{it}$ that is the product of the variables $QUINQ61\text{-}65_t$ and $OW_{it}$ and the variable $QUINQ61\text{-}65\_x\_RW_{it}$ that is the product of the variables $QUINQ61\text{-}65_{it}$ and $RW_i$. The variable $QUINQ61\text{-}65\_x\_OW_{it}$ is expected to have a negative effect while the variable $QUINQ61\text{-}65\_x\_RW_{it}$ is expected to have a positive one.

---

[14]The decision to give a value of 0.5 to the Middle East is discretionary, but due to the fact that this region includes Lebanon and Israel, two countries with non-Muslim majorities that have developed a wine industry.

When running regression (1) we need to take into account that exports' shares frequently take a value of zero and are also bounded between zero and one. For this type of regression Wagner (2001) suggests that a generalized linear model with a Logit as a link function, as proposed in Papke and Wooldridge (1996), is slightly more appropriate in theoretical terms than a Tobit estimation. However, given the popularity of the Tobit estimation in export studies and that the Papke and Wooldridge estimation procedure for the case of panel data has not yet made its way into any statistical software, the former is the one adopted in this paper.

## 4. Results

In Table 1 we see the results of regression (1). In column (1) we have the specification where all the measures of wine are done in volume while in column (2) they are in value.

**[Table 1 here]**

We can see that our hypothesis is supported by the data in both specifications. That is, an increase in the relative demand from the *RW* benefits the *NW* relatively to the *OW*. In specification (2) this is the case because the coefficients $\beta_1$, $\beta_4$ and $\beta_5$ are significant (at 1% the first two and 5% the latter) and $\beta_1 + \beta_5 > \beta_1 + \beta_4$. In specification (1) the coefficient $\beta_4$ is not significantly different from zero and so we again get that $\beta_1 + \beta_5 > \beta_1 + \beta_4$.

Two differences between the two specifications deserve special attention. The first one is that the variable *NIRW* has a negative and significant effect in the volume specification and a positive and significant one in the value specification. The second difference, which is the other side of the same coin, is that the variable *NIRW_x_NW (NIRW_ x_OW)* changes from having a positive and significant effect (being non-significant) in specification (1) to a negative and significant one in specification (2). However, given that we are controlling for many variables, no particular sign was expected for the variables *NIRW, NIRW_x_NW* and *NIRW_x_OW* by their own. Our model only predicts that the relationships discussed in the previous paragraph should hold. In any case, these seemingly unstable effects can be explained by the combination of two things. First, as is mentioned by the wine literature and commercial studies and has been highlighted in Villanueva (2011), some countries belonging to the *RW* are known to be re-exporters of wine, for example United Kingdom, United Arab Emirates, China/Hong Kong, to only mention a few. Second, it is also the case that the wine that is purchased with the aim of re-exportation is usually of a higher quality. The mentioned differences in sign and/or significance in the dummy variables *OW* and *NW* are likely being produced by the combination of these factors.

The variable *RMP* is non-significant in specification (1) and significant, but with the opposite to the expected sign in specification (2). The reason for this result is that many of the main wine exporters are countries located far from the importing ones; for instance, Australia, Argentina, Chile, South Africa and New Zealand in the Southern Hemisphere and the main new importing countries, for instance United Kingdom, Germany, Scandinavia and Japan in the Northern Hemisphere. In this respect we can say that the location of wine production and consequent exports was more determined by some other presumably omitted idiosyncratic factors rather than by our adopted measure of distance to potential markets.

The remaining of the control variables shows different levels of significance in the two specifications and do not deserve particular comments.

We now move to make one robustness test and subdivide the *NW* into two sub-worlds in order to see if our main hypothesis still holds. As it has been pointed out by Villanueva (2011) the *NW* is far from being a homogeneous group of producing and exporting countries. In particular, it can be subdivided into a Latin New World (*LNW*), composed by Argentina, Chile, Brazil, Mexico y Uruguay and an Anglo-Saxon New World (*ANW*), composed by United States, Australia, New Zealand, South Africa y Canada.

It is apparent that these two groups of countries have followed different production and commercial strategies. Even though the *LNW* was traditionally a wine producer, the *ANW* has been the one revolutionizing the production process, marketing and exports. This points out to the first difference, which is the late arrival of the *LNW* to the world´s wine export markets. This revolution was led by Australia in the beginning of the 1980's and followed up by the USA soon after and only adopted by the *LNW* in the beginning of the 1990's. Two of its main characteristics were a large scale of production as well as a simpler marketing strategy. Moreover, by around the beginning of the 1990's this process was complemented with a vigorous governmental support.

Even though the *LNW* imitated this commercial model, it still kept some of its traditional characteristics (i.e., smaller producers, more old-fashioned systems of vineyard management and wine-making as well as higher domestic per capita consumption). Finally, it should be stressed that the *ANW* emergence as a producer-exporter had from the very beginning a more export-oriented conception while the appearance of the *LNW* was a consequence of the shrinking of the domestic market.

As a consequence of the subdivision between *ANW* and *LNW* we will have the following new variables. $ANW_i$ is a dummy variable that takes the value 1 if the country belongs to the *ANW* and 0 otherwise and $LNW_i$ is a dummy variable that takes the value 1 if the country belongs to the *LNW* and 0 otherwise. The variable $NIRW\_x\_ANW_{it}$ is the product of the variables $NIRW_t$ and $ANW_i$ and indicates the effect that the *relative demand of the RW* has on the exports' share of each country belonging to the *ANW*. Similarly, the variable $NIRW\_x\_LNW_{it}$ is the product of the variables $NIRW_t$ and $LNW_i$ and indicates the effect that the *relative demand of the RW* has on the exports' share of each country belonging to the *LNW*.

In Table 2 we see the results of regression (1) for the case where the *NW* is divided into two sub-worlds. Again, in column (1) we have the specification for volume and in column (2) the one for values. We can clearly see that our hypothesis is robust to this subdivision of the new world. In both specifications of Table 2 the coefficients of the variables *NIRW x LNW* and *NIRW x ANW* are higher than the coefficient of the variable *NIRW x OW*; moreover, the latter is insignificant in specification (1). As expected, given the small size of the dataset, this subdivision reduces the significance level; in particular this is the case for the significance of the interactive effects *NIRW x LNW* and *NIRW x ANW*, in both specifications, relatively to the significance of *NIRW x NW* in Table 1.

**[Table 2 here]**

# 5. Conclusions

This paper uses a new available data and is the first historical and econometric study of the world wine exports' market in the period 1961 – 2005. It looks at what has determined the performance of the New World wine producers and how their success was favored by the relative increase in the demand for wine coming from countries with little tradition of wine consumption and practically inexistent tradition of wine production.

A merit of the paper is that it makes use of a combination of two separate literatures to explore a new phenomenon, which is the rapid appearance and success of the New World wine producers. Firstly, our paper is connected to the literature about the determinants of firms' exports propensity, which has developed a widely accepted econometric approach. Our study concentrates in the analysis of a particular industry, at country level and on a historical perspective. This marks a difference with this literature, which has mainly analyzed an aggregate of industries, at firm level and for only short periods of time.

Secondly, our paper brings a hypothesis from a sub-field of the Industrial Economics literature that analyzes oligopolistic competition in the presence of consumers' switching cost, where incumbent firms face the threat of entrance of new competitors. Our framework has some of the features that are typical in the switching cost model– i.e., it has traditional (or established) producers, new producers and new (unattached) buyers. The specific hypothesis that we tested is that the appearance of new (unattached) buyers favors the entrance of new competitors and / or those competitors with smaller export's share. As we already evidenced, the switching costs' empirical literature has not tested this hypothesis; hence this becomes our contribution to the switching cost empirical literature.

Finally, our paper puts forward the following future research agenda. The testing of the switching cost hypothesis can be extrapolated to other industries to see whether or not the appearance of new unattached buyers (if any) enhances the success of new producers. The feasibility of this extrapolation stems from the fact that the large expansion of exports that has occurred in the last half of the century was a worldwide phenomenon and not unique to the wine industry.

# References

Adelaide University, Monash University and the Australian Government (2007), *The global statistical compendium, 1961-2005*(Adelaide: Adelaide University).

Anderson, K. (2001), "The globalization and regionalization of wine"(Adelaide University Discussion Paper 0125).

Anderson, K., Norman, D. and Wittwer, G. (2003), *Globalization of the world's wine markets* (London: Blackwell Publishing Ltd).

Beggs, A. and Klemperer, P.D. (1989), "Multiperiod competition with switching costs"(Nuffield College, Oxford University Discussion Paper 45).

Borenstein, S. (1991), "Selling cost and switching cost: explaining retail gasoline margins", *RAND Journal of Economics*, 22(3), pp. 354-369.

Borenstein, S. and Rose, N. (1994), "Competition and price dispersion in the U.S. airline industry", *Journal of Political Economy*, 102(4), pp. 653-683.

Campbell, G. and Guibert, N. (2006), "Old World strategies against New World competition in a globalizing wine industry", *British Food Journal*, 108(4), pp. 233-242.

Central Intelligence Agency (2009), The C.I.A. world factbook (New York: Skyhorse Publishing).

Cholette, S., Castaldi, R. and Fredrick, A. (2005), "The globalization of the wine industry: implications for Old and New World producers" (San Francisco State University Discussion Paper).

Duncan, A. and Greenaway, D. (2008), "The Economics of Wine: Introduction", *The Economic Journal*, 118(529), pp. 137-141.

Farrell, J. (1986), "Moral hazard as an entry barrier", *RAND Journal of Economics*, 17(3), pp. 440-449.

Farrell, J. and Shapiro, C. (1988), "Dynamic competition with switching costs", *RAND Journal of Economics*, 19(1), pp. 123-137.

Farrell, J. and Klemperer, P. (2007), "Coordination and lock-in: competition with switching costs and network effects", *Handbook of Industrial Organization*, 3, pp. 1967-2072.

Gabszewicz, J., Pepall, L. and Thisse, J. (1992), "Sequential entry, with brand loyalty caused by consumer learning-by-doing-using", *Journal of Industrial Economics*, 40(4), pp. 397-416.

Green, R., Rodríguez Zúñiga, M. and Pierbattisti, L. (2003), "Global market changes and business behavior in the wine sector", (INRA-LORIA Discussion Paper 2).

Greenaway, D. and Kneller, R. (2007), "Firm heterogeneity, exporting and foreign direct investment", *Economic Journal*, 117, pp. 134-161.

Greenstein, S.M. (1993), "Did installed base give an incumbent any (measurable) advantage in federal computer procurement?", *RAND Journal of Economics*, 24(1), pp. 19-39.

Head, K. and Mayer, T. (2006), "Regional wage and employment responses to market potential in the EU", *Regional Science and Urban Economics*, 36(5), pp. 573-594.

Howard, J. and Sheth, J. (1969), The Theory of Buyer Behavior (New York: John Wiley & Sons).

Klemperer, P. (1987), "Entry deterrence in markets with consumer switching costs". *Economic Journal (Supplement),* 97, pp. 99-117.

Klemperer, P. (1995), "Competition when consumers have switching costs: an overview with applications to industrial organization, macroeconomics, and international trade", *The Review of Economic Studies*, 62(4), pp. 515-539.

Kneller, R., Pisu, M. and Yu, Z. (2008), "Overseas business costs and firm export performance", *Canadian Journal of Economics*, 41(2), pp. 639-669.

Lefebvre, E., Bourgault, M. and Lefebvre, L. A. (1998), "R&D related capabilities as determinants of export performance", *Small Business Economics*, 10(4), pp. 365-377.

Nassimbeni, G. (2001), "Technology, innovation capacity, and the export attitude of small manufacturing firms: a logit/tobit model", *Research Policy*, 30(2), pp. 365-377.

Papke, L.E. and Woolridge, J. M. (1996), "Econometric methods for fractional response variables with an application to 401(K) plan participation rates", *Journal of Applied Econometrics*, 11, pp. 619-632.

Ropern, S., Loven, J. H. and Añón Hıgón, D. (2006), "The determinants of export performance: evidence for manufacturing plants in Ireland and Northern Ireland", *Scottish Journal of Political Economy*, 53(5), pp. 586-615.

Schmalensee, R. (1982), "Product differentiation advantages of pioneering brands", *American Economic Review*, 72(3), pp. 349-365.

Seetharaman, P.B., Ainslie, A. and Chintagunta, P.K. (1999), "Investigating household state dependence effects across categories", *Journal of Marketing Research*, 36(4), pp. 448-500.

Serrano, R. and Pinilla, V. (2009), "Causes of world trade growth in agricultural and food products, 1951-2000: a demand function approach", *Applied Economics,* 42(27), pp. 3503-3518**.**

Simpson, J. (2009), "Old world versus New World: the origins of organizational diversity in the international wine industry, 1850-1914" (Universidad Carlos III Working Paper in Economic History 09-01).

Sterlacchini, A. (2001), "The determinants of export performance: a firm level study of Italian manufacturing", *Review of World Economics*, 137(3), pp. 450-472.

Shum, M. (2004), "Does advertising overcome brand loyalty? Evidence from the breakfast cereals market", *Journal of Economics & Management Strategy*, 13(2), pp. 241-272.

Van Dijk, M. (2002), "The determinants of export performance in developing countries: the case of Indonesian manufacturing" (Eindhoven Centre for Innovation Studies Working Paper 01).

Villanueva, E. (2011), *El boom exportador del Nuevo Mundo vitivinícola* (Barcelona: PhD Thesis, University of Barcelona).

Wagner, J. (1995), "Exports, firm size and firm dynamics", *Small Business Economics*, 7(1), pp. 29-39.

Wagner, J. (2001), "A note on the firm size export relationship", *Small Business Economics,* 17(4), pp. 229-237.

Wakelin, K. (1998), "Innovation and export behavior at the firm level", *Research Policy*, 26(7-8), pp. 829-841.

Wang, R. and Wen, Q. (1998), "Strategic invasion in markets with switching costs", *Journal of Economics and Management Strategy*, 7(4), pp. 521-549.

**Chart 1**

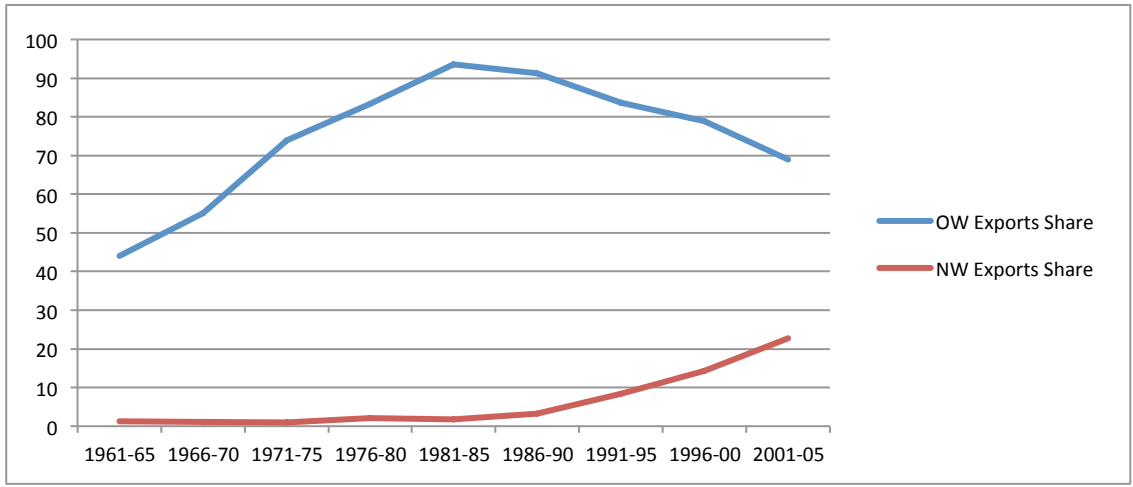

(a)

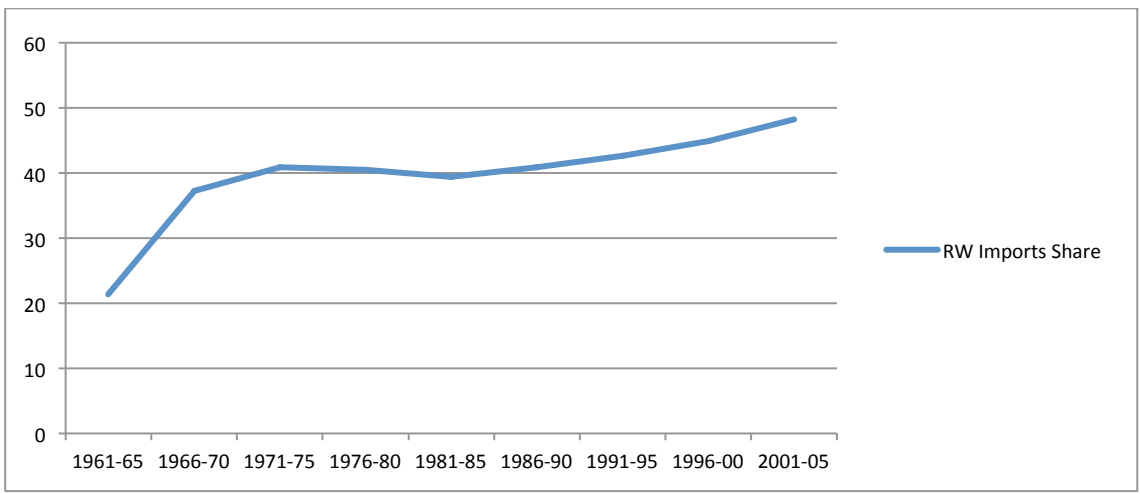

(b)

Note: OW exports' share = OW wine exports/worlds' total wine exports; NW exports' share = NW wine exports/worlds' total wine exports and RW imports' share = RW wine imports/worlds' total wine imports

# Chart 2

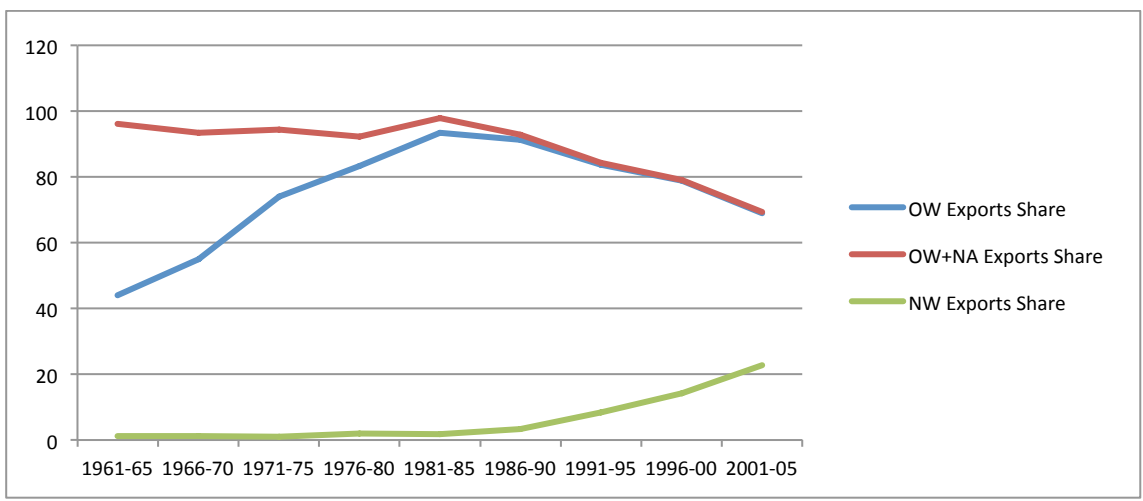

Note: OW exports' share = OW wine exports/worlds' total wine exports; OW+NA exports' share = OW+NA wine exports/worlds' total wine exports and NW exports' share = NW wine exports/worlds' total wine exports

# Table 1
# Regression (1) results

|  | (1) |  | (2) |  |
| --- | --- | --- | --- | --- |
| *NIRW* | -0.019* | (0.007) | 0.916* | (0.105) |
| *OW* | -0.066*** | (0.037) | 1.711* | (0.181) |
| *NW* | -0.129* | (0.040) | 0.048 | (0.135) |
| *NIRW x OW* | 0.029 | (0.021) | -3.111* | (0.328) |
| *NIRW x NW* | 0.056* | (0.018) | -0.572** | (0.231) |
| *RMP* | 0.000 | (0.000) | -0.003* | (0.000) |
| *EU 68-98* | 0.025* | (0.006) | -0.004 | (0.035) |
| *EURO 99-05* | 0.016*** | (0.009) | -0.039 | (0.051) |
| *CHRIST RULER* | 0.142* | (0.055) | 0.317* | (0.090) |
| *MUSLIM RULER* | -0.218* | (0.057) | -0.078 | (0.107) |
| *DIST LAT3050* | -0.005** | (0.002) | -0.009** | (0.004) |
| *AVERAGE TEMP* | 0.009* | (0.003) | 0.016* | (0.005) |
| *QUINQ61-65* | 0.003 | (0.008) | 0.087 | (0.067) |
| *QUINQ61-65_x_OW* | -0.014 | (0.017) | -0.583* | (0.143) |
| *QUINQ61-65_x_RW* | -0.009 | (0.009) | 0.246* | (0.073) |

Note: Standard errors in parentheses, *** indicates significance at 10%, ** at 5%, and * at 1%. Specification (1) is in volume and specification (2) is in value.

# Table 2

# Regression results when New World is sub-divided into Latin and Anglo-Saxon New Wine Worlds

| | (1) | | (2) | |
|---|---|---|---|---|
| *NIRW* | -0.019* | (0.007) | 0.914* | (0.106) |
| *OW* | -0.066*** | (0.038) | 1.709* | (0.181) |
| *LNW* | -0.140* | (0.052) | 0.019 | (0.166) |
| *ANW* | -0.120** | (0.049) | 0.069 | (0.152) |
| *NIRW x OW* | 0.029 | (0.021) | -3.108* | (0.328) |
| *NIRW x LNW* | 0.054** | (0.022) | -0.530*** | (0.282) |
| *NIRW x ANW* | 0.058* | (0.020) | -0.601** | (0.257) |
| *RMP* | 0.000 | (0.000) | -0.003* | (0.000) |
| *EU 68-98* | 0.025* | (0.006) | -0.004 | (0.035) |
| *EURO 99-05* | 0.016*** | (0.010) | -0.039 | (0.051) |
| *CHRIST RULER* | 0.145* | (0.056) | 0.318* | (0.091) |
| *MUSLIM RULER* | -0.216* | (0.057) | -0.077 | (0.107) |
| *DIST LAT3050* | -0.005** | (0.002) | -0.009** | (0.004) |
| *AVERAGE TEMP* | 0.010* | (0.003) | 0.016* | (0.006) |
| *QUINQ61-65* | 0.002 | (0.008) | 0.086 | (0.068) |
| *QUINQ61-65_x_VM* | -0.014 | (0.017) | -0.582* | (0.144) |
| *QUINQ61-65_x_RM* | -0.009 | (0.009) | 0.245* | (0.074) |

Note: Standard errors in parentheses, *** indicates significance at 10%, ** at 5%, and * at 1%. Specification (1) is in volume and specification (2) is in value.